\documentclass[aps,prd,showpacs,nofootinbib,preprint]{revtex4}

\usepackage{amsmath}
\usepackage{amssymb}
\usepackage{graphicx}

\begin{document}
\pagestyle{plain}

\title{Reheating the Universe in Braneworld Cosmological Models with bulk-brane energy transfer}
\author{T. Harko}
\email{harko@hkucc.hku.hk}
\affiliation{Department of Physics and
Center for Theoretical and Computational Physics, University of
Hong Kong, Pok Fu Lam Road, Hong Kong, P. R. China}
\author{W. F. Choi}
\affiliation{Department of Physics
and Center for Theoretical and Computational Physics, University
of Hong Kong, Pok Fu Lam Road, Hong Kong, P. R. China}
\author{K. C. Wong}
\affiliation{Department of Physics
and Center for Theoretical and Computational Physics, University
of Hong Kong, Pok Fu Lam Road, Hong Kong, P. R. China}
\author{K. S. Cheng}
\affiliation{Department of Physics
and Center for Theoretical and Computational Physics, University
of Hong Kong, Pok Fu Lam Road, Hong Kong, P. R. China}

\date{\today}

\begin{abstract}

The emergence of the cosmological composition (the reheating era)
after the inflationary period  is analyzed in the framework of the
braneworld models, in which our Universe is a three-brane embedded
in a five-dimensional bulk, by assuming the possibility of the
brane-bulk energy exchange. The inflaton field is assumed to decay
into normal matter only, while the dark matter is injected into
the brane from the bulk. To describe the reheating process we
adopt a phenomenological approach, by describing the decay of the
inflaton field by a friction term proportional to the energy
density of the field. After the radiation dominated epoch the
model reduces to the standard four dimensional cosmological model.
The modified field equations are analyzed analytically and
numerically in both the extra-dimensions dominate reheating phase
(when the quadratic terms in energy density dominate the
dynamics), and in the general case. The evolution profiles of the
matter, of the scalar field and of the scale factor of the
universe are obtained for different values of the parameters of
the model, and of the equations of state of the normal and dark
matter, respectively. The equation describing the time evolution
of the ratio of the energy density of the dark and of the normal
matter is also obtained. The ratio depends on the rate of the
energy flow between the bulk and the brane. The observational
constraint of an approximately constant ratio of the dark and of
the baryonic matter requires that the dark matter must be
non-relativistic (cold). The model predicts a reheating
temperature of the order of $3\times 10^6$ GeV, a brane tension of
the order of $10^{25}$ GeV$^4$, and the obtained composition of
the universe is consistent with the observational data.
\end{abstract}

 \maketitle

\section{Introduction}

Early studies on superstring theory and M-theory have suggested
that our four dimensional world is embedded into a higher
dimensional spacetime. Particularly, the 10 dimensional $E_8
\otimes E_8$ heterotic superstring theory is a low-energy limit of
the 11 dimensional supergravity, under the compactification scheme
$M^{10}\times S_1 / Z_2$ \cite{Witten,Witten2}. Thus, the 10
dimensional spacetime is compactified as $M^4 \times CY^6 \times
S_1 / Z_2$, implying that our universe (a brane) is embedded into
a higher dimensional bulk. In this paradigm, the standard model
particles are open strings, confined on the braneworld, whilst the
gravitons and the closed strings can freely propagate into the
bulk \cite{Polchinski}.

Among the collection of the braneworld scenarios, the
Randall-Sundrum Type II model has the virtue of providing a new
type of compactification of gravity \cite{Randall,Randall2}.
Standard 4D gravity can be recovered in the low-energy limit of
the model, with a 3-brane of positive tension embedded in 5D
anti-de Sitter bulk. The covariant formulation of the brane world
models has been formulated in \cite{Shiromizu}, leading to the
modification of the standard Friedmann equations on the brane. It
turns out that the dynamics of the early universe is altered by
the quadratic terms in the energy density and by the contribution
of the components of the bulk Weyl tensor, which both give a
contribution in the energy momentum tensor. This implies a
modification of the basic equations describing the cosmological
and astrophysical dynamics, which has been extensively considered
recently \cite{all2}.

The recent observations of the CMB anisotropy by WMAP
\cite{Spergel} have provided convincing evidence for the
inflationary paradigm \cite{Guth}, according to which in its very
early stages the universe experienced an accelerated (de Sitter)
expansionary phase (for recent reviews on inflation see
\cite{infl}).

At the end of inflation,  the universe is in a cold and
low-entropy phase, which is utterly different from the present hot
high-entropy universe. Therefore the  universe should be reheated,
or defrosted, to a high enough temperature, in order to recover
the standard Hot Big Bang \cite{reh}. The reheating process may be
envisioned as follows: the energy density in zero-momentum mode of
the scalar field decays into normal particles with decay rate
$\Gamma$. The decay products then scattered and thermalize to form
a plasma \cite{infl}.

Apart from the behavior of the inflaton field, the evolutions of
dark energy and dark matter in reheating stage were also
considered. In \cite{Susperregi}, dark energy and dark matter were
originated from a scalar field in different stages of the
inflation, according to a special form of potential. Meanwhile,
the conditions for unifying the description of inflation, dark
matter and dark energy were considered in \cite{Liddle3}. A
specific model was later proposed in \cite{Cardenas}, by using a
modified quadratic scalar potential. The candidates of dark matter
in \cite{Liddle3} and \cite{Cardenas} were oscillations of a
scalar field. However, it may be possible that dark matter existed
on its own without originating from the scalar field. This may
pose less stringent constraint on the scalar field, so that dark
matter can be included in inflation paradigm in a easier way. On
the other hand, it was proposed that the decay products of scalar
field acquired thermal mass \cite{Kolb}.

The reheating in the braneworld models has also been considered
recently. In the context of the braneworld inflation driven by a
bulk scalar field, the energy dissipation from the bulk scalar
field into the matter on the brane was studied in \cite{HiTa03}.
The obtained results supports the idea that the brane inflation
model, caused by a bulk scalar field, may be a viable alternative
scenario of the early universe.  The inflation and reheating in a
brane world model derived from Type IIA string theory was studied
in \cite{BrDa03}. In this model the inflaton can decay into scalar
and spinor particles, thus reheating the universe.  A model in
which high energy brane corrections allow a single scalar field to
describe inflation at early epochs and quintessence at late times
was discussed in \cite{SaDaSh03}. The reheating mechanism in the
model originates from Born-Infeld matter, whose energy density
mimics cosmological constant at very early times and manifests
itself as radiation subsequently.  The particle production at the
collision of two domain walls in a 5-dimensional Minkowski
spacetime was studied in \cite{TaMa04}. This may provide the
reheating mechanism of an ekpyrotic (or cyclic) brane universe, in
which two BPS branes collide and evolve into a hot big bang
universe. The reheating temperature $T_{RH}$ in models in which
the universe exits reheating at temperatures in the MeV regime was
studied in \cite{Hannestad}, and a minimum bound on $T_{RH}$ was
obtained. The derived lower bound on the reheating temperature
also leads to very stringent bounds on the compactification scale
in models with $n$ large extra dimensions.  The dark matter
problem in the Randall-Sundrum type II brane world scenario was
discussed in \cite{Pa05}, by assuming that the lightest
supersymmetric particle is the axino. The axinos can play the role
of cold dark matter in the universe, due to the higher reheating
temperatures in the brane world model, as compared to the
conventional four-dimensional cosmology. The impact of the
non-conventional brane cosmology on the relic abundance of
non-relativistic stable particles in high and low reheating
scenarios was investigated in \cite{DaKh06}. In the case of high
reheating temperatures, the brane cosmology may enhance the dark
matter relic density by many order of magnitudes, and a stringent
lower bound on the five dimensional scale may be obtained. In the
non-equilibrium case, the resulting relic density is very small.
The curvaton dynamics in brane-world cosmologies was studied in
\cite{PaZa06}. By assuming that the inflaton field survives
without decay after the end of inflation, the curvaton reheating
mechanism was applied to the Randall-Sundrum model and to its
curvature corrections.

Recently, motivated by particle physics results, the possibility
of an energy exchange between the bulk and the brane was also
suggested.  The observational constraints on the cosmological
models in the brane-world scenario in which the bulk is not empty,
and exchange of mass-energy between the bulk and the bane is
allowed, were explored in \cite{Umezu}. The evolution of matter
fields on the brane is modified due to new terms in the energy
momentum tensor describing this exchange. This model accounts for
the observed suppression of the CMB power spectrum at low
multipoles, and the observed recent cosmic acceleration is
attributable to the flow of matter from the bulk to the brane. The
cosmological evolution of a brane with a general bulk matter
content was considered in \cite{BoTa06}, by assuming that the bulk
pressure and the energy exchange densities are comparable to the
brane energy density. By adopting a phenomenological fluid ansatz
and generalizations of it, a set of exact solutions of the
Friedmann equation that exhibit accelerated expansion were
derived.  The braneworld models with bulk-brane energy exchange
allow for crossing of the $w=-1$ phantom divide line without
introducing phantom energy with quantum instabilities
\cite{Bogdanos}. The class of the braneworld models with
bulk-brane energy exchange have the ability to provide crossing of
the phantom divide, and the observational cosmological data hint
towards natural values for the model parameters.

It is the purpose of the present paper to construct a model of
braneworld reheating by adopting as a key feature the energy
exchange between the bulk and the brane during the reheating
stage. The era of focus will be from the end of inflation until
the present time, in contrast with \cite{Umezu} and
\cite{BoTa06,Bogdanos}. In order to explain the present day
composition of our universe, we adopt, as an essential
characteristic of our model, the hypothesis that the existence of
the dark matter is the result of the real matter exchange between
the bulk and the brane. However, the other matter fields bare no
such freedom, according to the confinement conditions introduced
in \cite{Polchinski}.

The main physical features of our model can be presented as
follows: the dark matter from the bulk and flows to the brane
after the inflationary phase, while the scalar fields and the
normal matter are always confined on the brane. Therefore the bulk
leaves an imprint on the brane by injecting dark matter. The
effect of dark matter flowing into the brane is studied in the
context of the cosmological composition, with the reheating
temperature as an important parameter.  The inflaton field decays
into the normal matter field \cite{reh,Harko}, attaining thermal
equilibrium and acquiring plasma masses through the mechanism
proposed in \cite{Kolb}. Therefore, as reheating is concerned, the
presence of the scalar field is essential, and it decays into
visible matter only. Furthermore, the specific forms of the
inflaton potential are not essential in our approach, since during
reheating, the kinetic term $\dot{\phi}^2$ plays a crucial role.
The main focus of the present paper is to study the dynamics of
the universe by taking into account the extra-dimensional effects
on the reheating process. The ratio $u$ of the energy densities of
the dark matter and of the normal matter is an important parameter
of our model, and its time evolution is studied by using both
analytical and numerical methods. As a test of the model, several
important observational parameters like the maximum density of the
matter in the universe and the reheating temperature $T_{reh}$ are
also estimated. Hence, we adopt as  our main observational
constraints the constancy of the ratio $u$ after the photon
decoupling, the absence of the inflaton in the present day
universe, and the bound of $T_{reh}$ from gravitino product,
\cite{Ellis,Kawasaki,Moroi}.


The present paper is organized as follows.  The basic equations of
our model are formulated in Section II. A dimensionless form of
the field equations is presented in Section III, where the
physically acceptable range of the values of the physical
parameters is also discussed. The extra-dimensional effects
dominated phase of the reheating, during which the quadratic terms
in the energy density dominate the dynamics and evolution of the
universe, is discussed in Section IV. The reheating process is
considered in its full generality in Section V, and the role of
the linear terms in the energy density is investigated. Finally,
in Sec. VI we summarize and conclude our results.

\section{Geometry and field equations in the bulk-brane energy transfer model}

We start by considering a five dimensional ($5D$) spacetime (the
bulk), with a single four-dimensional ($4D$) brane, on which usual
(baryonic) matter and physical fields are confined. The $4D$ brane
world $({}^{(4)}M,g_{\mu \nu })$ is located at a hypersurface
$\left(B\left( X^{A}\right) =0\right)$ in the $5D$ bulk spacetime
$({}^{(5)}M,g_{AB})$, of which coordinates are described by
$X^{A},A=0,1,...,4$. The induced $4D$ coordinates on the brane are
$x^{\mu },\mu =0,1,2,3$.

The action of the system is given by~\cite{Shiromizu}
\begin{equation}
S=S_{bulk}+S_{brane},  \label{bulk}
\end{equation}
where
\begin{equation}
S_{bulk}=\int_{{}^{(5)}M}\sqrt{-{}^{(5)}g}
\left[\frac{1}{2k_{5}^{2}}{}^{(5)}R+{}^{(5)}L_{m}+\Lambda_{5}\right]
d^{5}X,
\end{equation}
and
\begin{equation}
S_{brane}=\int_{{}^{(4)}M}\sqrt{-{}^{(5)}g}
\left[\frac{1}{k_{5}^{2}}K^{\pm}+L_{brane}\left( g_{\alpha \beta
},\phi \right)+ \lambda _{}\right] d^{4}x,
\end{equation}
where $k_{5}^{2}=8\pi G_{5}$ is the $5D$ gravitational constant,
${}^{(5)}R$ and ${}^{(5)}L_{m}$ are the $5D$ scalar curvature and
the matter Lagrangian in the bulk, $L_{brane}\left( g_{\alpha
\beta },\phi \right)$ is the 4d Lagrangian, which is given by a
generic functional of the brane metric $g_{\alpha \beta }$ and of
the matter fields $\phi $, $K^{\pm }$ is the trace of the
extrinsic curvature on either side of the brane, and $\Lambda
_{5}$ and $\lambda $ (the constant brane tension) are the negative
vacuum energy densities in the bulk and on the brane,
respectively. The brane tension is the characteristic energy scale
of the brane, below which the quadratic terms in the field
equations, derived from the action  Eq.~(\ref{bulk}), can be
neglected. Thus, in the limit $\lambda \rightarrow \infty$ the
field equations reduce to the usual form of the standard general
relativity.

The Einstein field equations in the bulk are given
by~\cite{Shiromizu}
\begin{equation}
{}^{(5)}G_{IJ}=k_{5}^{2} {}^{(5)}T_{IJ},\qquad
{}^{(5)}T_{IJ}=-\Lambda _{5}
{}^{(5)}g_{IJ}+\delta(B)\left[-\lambda_{}
{}^{(5)}g_{IJ}+T_{IJ}\right] ,
\end{equation}
where
\begin{equation}
{}^{(5)}T_{IJ}\equiv - 2\frac{\delta {}^{(5)}L_{m}}{\delta
{}^{(5)}g^{IJ}} +{}^{(5)}g_{IJ} {}^{(5)}L_{m},
\end{equation}
is the energy-momentum tensor of bulk matter fields, while $T_{\mu
\nu }$ is the energy-momentum tensor localized on the brane and
which is defined by
\begin{equation}
T_{\mu \nu }\equiv -2\frac{\delta L_{brane}}{\delta g^{\mu \nu
}}+g_{\mu \nu }\text{ }L_{brane}.
\end{equation}

The delta function $\delta \left( B\right) $ denotes the
localization of brane contribution. In the $5D$ spacetime a brane
is a fixed point of the $Z_{2}$ symmetry. In the following we
assume
\begin{equation}
{}^{(5)}L_{m}\neq 0.
\end{equation}

The induced $4D$ metric is $g_{IJ}={}^{(5)}g_{IJ}-n_{I}n_{J}$,
where $n_{I}$ is the space-like unit vector field normal to the
brane hypersurface ${}^{(4)}M$. The basic equations on the brane
are obtained by projections onto the brane world, and are given by
\cite{Ap}
\begin{equation}
G_{\mu \nu }=-3\lambda g_{\mu \nu }+\frac{\lambda
^{(5)}}{24k_{5}^{4}}T_{\mu
\nu }+\frac{1}{4k_{5}^{4}}S_{\mu \nu }-\varepsilon _{\mu \nu }+\frac{1}{%
3k_{5}^{4}}F_{\mu \nu },
\end{equation}
where
\begin{equation}
S_{\mu \nu }=\frac{1}{2}TT_{\mu \nu }-\frac{1}{4}T_{\mu \alpha
}T_{\nu }^{\alpha }+\frac{3T_{\alpha \beta }T^{\alpha \beta
}-T^{2}}{24}g_{\mu \nu },
\end{equation}
\begin{equation}
\varepsilon _{\mu \nu }=C_{ABCD}n^{C}n^{D}g_{\mu }^{A}g_{\nu
}^{B},
\end{equation}
and
\begin{equation}
F_{\mu \nu }=^{(5)}T_{AB}g_{\mu }^{A}g_{\nu }^{B}+\left(
^{(5)}T_{AB}n^{A}n^{B}-\frac{1}{4}^{(5)}T\right) g_{\mu \nu },
\end{equation}
respectively.

Apart from the terms quadratic in the brane energy-momentum
tensor, in the field equations on the brane there are two
supplementary terms, corresponding to the projection of the $5D$
Weyl tensor $\varepsilon _{\mu \nu }$ and of the projected tensor
$F_{\mu \nu }$, which contains the bulk matter contribution. Both
terms induce bulk effects on the brane.

For cosmological applications we adopt a metric of the form
\begin{equation}
ds^{2}=-n^{2}(t,\xi )dt^{2}+a^{2}\left( t,\xi \right) \gamma _{\mu
\nu }dx^{\mu }dx^{\nu }+d\xi ^{2},
\end{equation}
where $\gamma _{\mu \nu }$ is the maximally symmetric 3-space, and
we denoted the fifth coordinate by $x^5=\xi $. $a$ and $n$ are the
scale and shift factors, respectively. The Hubble parameter $H$ on
the brane, describing the cosmological dynamics of the universe,
is defined as $H=\dot{a}/a$. For further discussions on the
gravitational field equations in the brane world models and their
cosmological applications see
\cite{Kiritsis,Tetradis,Apostolopoulos,Myung,Tetradis2,Apostolopoulos2}.

On the brane the matter, consisting of normal matter, dark matter,
a scalar field etc., is in a perfect fluid form, and is described
in terms of a total energy density $\rho _{total}=\sum_{i=1}^n\rho
_i$, where $\rho _i$, $i=1,...,n$ are the energy densities of the
individual components, and a total thermodynamic pressure
$p_{total}=\sum_{i=1}^np _i$, where $p_i$, $i=1,...,n$ are the
pressures of the matter fluid components. The corresponding
energy-momentum tensor on the brane is
\begin{equation}
T^{\mu}_{\nu}= \delta(\xi ){\rm diag}\left(-\rho
_{total},p_{total},p_{total},p_{total}\right).
\end{equation}

The delta function $\delta(\xi )$ implies the confinement of the
matter fields on the brane at $\xi =0$.


We assume that the matter component in the bulk is in the form of
dark matter, with energy density $\rho _{DM}$ and pressure
$p_{DM}$, respectively, with energy-momentum tensor
\begin{equation}
T^A_B=\left(\rho _{DM}+p_{DM}\right)U^AU_B+p_{DM}\delta ^A_B.
\end{equation}

Then the energy momentum tensor $T^{A}_{B}$ of dark matter in the
bulk has the non-zero components :
\begin{equation}
T^0_5 \sim (\rho _{DM}+p_{DM}) U_5      \,\,\,\, , \,\,\,\, T^5_5=
(\rho _{DM}+p_{DM})U^5 U_5 + p_{DM}, \label{T-0,5}
\end{equation}
where the five-velocity $U_5$ represents the matter flow from the
bulk to the brane \cite{Umezu}. Hence in our model the dark matter
can exist simultaneously in the bulk and on the brane.

The term $2T^0_5$ is the discontinuity of the (0,5) component of
energy momentum tensor at $\xi =0$. In the static bulk with
respect to the expanding brane, $U_5$ is given by $U_5 \propto -lH
$, where $l=\sqrt{-6M_5^3/\Lambda _5}$ is the bulk curvature
radius, and $M_5$ and $\Lambda_5$ are the Planck mass and the
cosmological constant in the 5 dimensional bulk, respectively
\cite{Mukohyama2,Ichiki}. We assume that the dark matter obeys an
equation of state of the form
\begin{equation}
p_{DM}=(w-1)\rho _{DM},
\end{equation}
where $1\leq w\leq 2$ is a constant.

To obtain $T^0_5$ we follow the approach developed in
\cite{Umezu}. Since
$\rho _{DM}=\rho_{cr}\left(a_0/a\right)^{3w}$, where $\rho_{cr}$
is the present day critical density and $a_0$ is the value of the
scale factor at the moment when the energy transfer from the bulk
to the brane begins, we obtain
\begin{equation}
T^0_5=-\frac{\alpha_{bb}}{2}\rho_{cr}\left(\frac{a_0}{a}\right)^{3w}H.
\end{equation}

The dimensionless parameter $\alpha_{bb}$ describes the flow of
dark matter from bulk to brane. Finally, the evolution equation of
the dark matter on the brane is \cite{Umezu}
\begin{equation}
\dot{\rho }_{DM} + 3H (\rho_{DM} + p_{DM} )= - 2T^0_5
\end{equation}

In the widely accepted inflationary scenario it is assumed that
during an initial period the universe is dominated by a large,
approximately constant potential term $V(\phi )$ of a scalar field
$\phi $, known as the inflaton field \cite{Guth,infl}. The
energy-momentum tensor of the scalar field can be written in the
perfect fluid form, with energy density $\rho _{\phi }$ and
pressure $p_{\phi }$ given by
\begin{equation}
\rho_{\phi}=\frac{\dot{\phi}^2}{2} + V(\phi ),
\end{equation}
and
\begin{equation}
p_{\phi}=\frac{\dot{\phi}^2}{2} - V(\phi ),
\end{equation}
respectively. During a second period of evolution, the potential
minimum is approached, $V(\phi )$ tends to zero, and the scalar
field starts to fluctuate violently around the minimum value. The
coupling to the field component becomes effective and the scalar
field energy is converted into matter energy, thus giving rise to
a large increase of the entropy \cite{infl,reh}.

Inflationary scenarios have always to face the problem of the
transition from a de Sitter stage, during which the evolution of
the universe is dominated by the scalar field, to a subsequent
radiation - or matter - dominated Friedmann-Robertson-Walker type
cosmological model. One of the possible approaches to this problem
is phenomenological \cite{infl,reh,Harko}, and consists in the
introduction of a suitable loss term in the scalar field dynamics,
which also appears as a source term for the energy density of the
matter fluid. It is this source term which is responsible for the
reheating process that follows the adiabatic supercooling process
during the de Sitter phase. To describe mathematically the process
is convenient to introduce a "friction" term $\Gamma $, describing
the decay rate of the scalar field, and a corresponding term for
the matter fluid.

Therefore, the basic equations describing the reheating phase
after braneworld inflation, with matter exchange between the brane
and the bulk, are given by
\begin{equation}
H^2=\left(\frac{\dot{a}}{a}\right)^2=\frac{8 \pi G
\rho_{total}}{3}\left(1+\frac{\rho_{total}}{2\lambda}\right)
-\frac{k}{a^2} +\Lambda_4+\chi ,   \label{eq1}
\end{equation}
\begin{equation}
\dot{\rho }_m+3H(\rho_m+p_m)= \Gamma \left(\rho _m,\rho _{\phi
}\right)\rho _{\phi },
\end{equation}
\begin{equation}
\dot{\rho} _\phi +3H(\rho_\phi+p_\phi)= -\Gamma \left(\rho _m,\rho
_{\phi }\right)\rho _{\phi },
\end{equation}
\begin{equation}
\dot{\rho} _{DM} + 3H (\rho_{DM} + p_{DM} )=
\frac{\alpha_{bb}}{2}\rho_{cr}\left(\frac{a_0}{a}\right)^{3w}H,
\label{eq:DM evolution}
\end{equation}
\begin{equation}
\dot{\chi }+4\frac{\dot{a}}{a}\left(\chi
+\frac{k_5^2}{6}T_5^5\right)=\frac{k_5^4}{9}\left(\rho
_{total}+\lambda \right)T_5^0,
\end{equation}
\begin{equation}\label{eqd}
\rho_{total}=\rho_m + \rho_\phi + \rho_{DM},
\end{equation}
where $\Lambda_4$ is the cosmological constant on the brane, $\chi
$ is the so-called dark radiation term, and we have assumed the
existence of a single matter component on the brane, with energy
density $\rho _m$ and pressure $p_m$, respectively.

\section{Dimensionless form of the field equations}    \label{Physical Assumptions}

The system of equations Eqs.~(\ref{eq1})-(\ref{eqd}) can be
simplified by adopting some, physically justified, approximations.
First of all, the $4D$ cosmological constant $\Lambda _4$ can be
neglected, as its derived value from the 3rd year WMAP data
 is around $10^{-47}${\rm GeV}$^4$ \cite{Spergel}, which is
a small fraction of the energy density up to at least the photon
decoupling period. Also the curvature term can be taken as zero,
$k=0$, as the universe is observed to be nearly flat. Moreover,
under the conditions considered in this paper, the dark radiation
term
\begin{equation}\label{expl1}
\chi \propto \frac{1}{a^4}
\end{equation}
is negligibly small as compared with the other components
\cite{Umezu}.

An important relation to characterize the matter is the equation
of state (EOS) $p_m=\left(\gamma -1\right)\rho _m$. The product
particles have different equations of state in various time
scales. The index $\gamma $ is $\gamma =4/3$ for radiation before
the radiation-matter equality time at $t_{eq}= 10^5$years $\approx
10^{11}$ sec., and $\gamma \approx 1$ for non-relativistic matter
after $t_{eq}$.  The constant $w$ determining the EOS of dark
matter will be fixed in various scenarios. It takes the value
$w=4/3$ for hot dark matter (HDM), and $w\approx 1$ for cold dark
matter (CDM), just like in the case of ordinary matter.

Next we consider the scalar field properties. In the
post-inflationary stage, the inflaton field executes coherent
oscillations about the minimum of the potential \cite{Kolb2}.
Therefore the kinetic term dominates the potential term in the
reheating era, thus leading to
\begin{equation}
\rho_{\phi}=\frac{\dot{\phi}^2}{2} + V(\phi ) \approx
\frac{\dot{\phi}^2}{2} \approx p_{\phi}.
\end{equation}

The explicit expression of the decay width of the scalar field can
be represented as \cite{Kolb},
\begin{equation}
\Gamma= \alpha_{\phi} M_{\phi}
\sqrt{1-\left(\frac{T}{M_{\phi}}\right)^2}\label{decay width}
\end{equation}
where $\alpha_{\phi}$ and $M_{\phi}$ are the coupling constant and
the mass of inflaton respectively. Assuming that the thermodynamic
equilibrium is established, $T$ is the temperature of the decay
product, and it can be related to the matter density $\rho_m$.
Generally,
\begin{equation}\label{expl4}
\rho_m=\sigma T^{\gamma /\left(\gamma -1\right)},
\end{equation}
where $\sigma $ is a constant.  If the matter is in the form of
radiation, $\rho _m=\pi ^2 T^4/15$, and $\sigma =\pi ^2/15$. In
the matter-dominated universe, the relation between $\rho_m$ and
$T$ can also be written down explicitly.  The scalar field is
negligible small in the matter-dominated phase. In the
non-relativistic matter domination era $T \leq 1 {\rm eV}$. This
is far smaller than the minimum bound obtained for $M_\phi $
\cite{Kolb}. Hence, in the non-relativistic phases of matter
evolution the decay rate is simply $\Gamma =\alpha_\phi M_\phi$.

Thus, by defining the parameter
\begin{equation}
\kappa^2=\frac{4\pi G}{3\lambda},
\end{equation}
and with the use of Eqs. (\ref{expl1}) - (\ref{expl4}), the system
of equations describing the reheating process on the brane in the
presence of bulk-brane energy exchange is given by
\begin{equation}
\frac{1}{a}\frac{da}{dt}=H=\left(\frac{8\pi G}{3} \rho_{total}+
\kappa^2 \rho_{total}^2\right)^\frac{1}{2}
\end{equation}
\begin{equation}
\dot{\rho }_{m}= - 3\gamma H \rho_m + 2 \alpha_{\phi} M_{\phi}
\rho_{\phi} \sqrt{1-\frac{1}{M_{\phi }^2}\left(\frac{\rho
_m}{\sigma }\right)^{2\left(\gamma -1\right)/\gamma }},
\end{equation}
\begin{equation}
\dot{\rho }_{\phi } = - 6 H \rho _\phi -2 \alpha_{\phi} M_{\phi}
\rho_{\phi} \sqrt{1-\frac{1}{M_{\phi }^2}\left(\frac{\rho
_m}{\sigma }\right)^{2\left(\gamma -1\right)/\gamma }},
\end{equation}
\begin{equation}
\dot{\rho }_{DM} = H
\left[\frac{\alpha_{bb}}{2}\rho_{cr}\left(\frac{a_0}{a}\right)^{3w}H
- 3w\rho _{DM}\right],
\end{equation}

In order to simplify the study of the evolution equations during
the reheating period, we rescale the time variable and the
densities, by introducing a set of dimensionless variables $\left(
\tau ,r_{m},r_{DM},r_{\phi },r_{cr}\right) $, defined as
\begin{equation}\label{trans1}
t=B\tau ,\rho _{m}=Ar_{m},\rho _{DM}=Ar_{DM},\rho _{\phi
}=Ar_{\phi },\rho _{cr}=Ar_{cr},
\end{equation}
where
\begin{equation}\label{trans2}
B=\frac{1}{2\alpha _{\phi }M_{\phi }},
\end{equation}
and
\begin{equation}\label{trans3}
A=\frac{12\alpha _{\phi }^{2}M_{\phi }^{2}}{8\pi G},
\end{equation}
respectively. By denoting
\begin{equation}
\eta ^{2}=\frac{6\alpha _{\phi }^{2}M_{\phi }^{2}}{8\pi G\lambda },\omega =%
\frac{1}{M_{\phi }^{2}}\frac{\left( \sqrt{12}\alpha _{\phi
}M_{\phi }\right) ^{4\left( \gamma -1\right) /\gamma }}{\left(
8\pi G\sigma \right) ^{2\left( \gamma -1\right) /\gamma
}},\varepsilon =\alpha _{bb}r_{cr},
\end{equation}
the system of equations describing the reheating phase with
bulk-brane energy exchange is given by
\begin{equation}
\frac{1}{a}\frac{da}{d\tau }=h(\tau )=\sqrt{r_{m}+r_{DM}+r_{\phi
}+\eta ^{2}\left( r_{m}+r_{DM}+r_{\phi }\right) ^{2}},  \label{f1}
\end{equation}
\begin{equation}
\frac{dr_{m}}{d\tau }=-3\gamma h(\tau )r_{m}+r_{\phi
}\sqrt{1-\omega r_{m}^{2\left( \gamma -1\right) /\gamma }},
\label{f3}
\end{equation}
\begin{equation}
\frac{dr_{\phi }}{d\tau }=-6h\left( \tau \right) r_{\phi }-r_{\phi }\sqrt{%
1-\omega r_{m}^{2\left( \gamma -1\right) /\gamma }}.  \label{f4}
\end{equation}
\begin{equation}
\frac{dr_{DM}}{d\tau }=h(\tau )\left[ \varepsilon \left( \frac{a_{0}}{a}%
\right) ^{3w}-3wr_{DM}\right] ,  \label{f2}
\end{equation}

The system of equations Eqs. (\ref{f1})-(\ref{f4}) must be
considered together with the initial conditions
\begin{equation}
a(0)=a_{0},r_{DM}(0)=0,r_{m}(0)=0,r_{\phi }(0)=r_{\phi 0},
\end{equation}
where $a_{0}$ and $r_{\phi 0}$ are the initial values of the scale
factor and of the scalar field energy density at the beginning of
the reheating phase.

In order to obtain a consistent physical interpretation of our
results, we need to know the numerical values of the model
parameters, which are $\alpha _{\phi }$, $M_{\phi }$ and $\lambda
$, respectively. $\alpha_\phi$ and $M_\phi$ are not known
separately, but the whole quantity $\alpha_\phi M_\phi$ can be
estimated. From Eq.(\ref{decay width}), it follows that the order
of magnitude of $\alpha_\phi M_\phi$ is about the order of
magnitude of the decay width, which is the reciprocal of the
characteristic timescale of reheating. The inflationary era ends,
and reheating can start at the earliest at around $t= 10^{-32}$ s,
while the Hot Big Bang commences at around $t= 10^{-18}$ s
\cite{Liddle2,Kolb2}. The reheating process should complete before
the Hot Big Bang to restore the BBN. Therefore $10^{18}$
sec.$^{-1} \leq \Gamma \leq 10^{32}$ sec.$^{-1}$ implies that
\begin{equation}
1\,\mbox{keV}  \leq  \alpha_\phi M_\phi  \leq   10^8\,\mbox{GeV}.
\end{equation}

As for the brane tension $\lambda$, it is constrained by the
validity of Newtonian gravity in 4 dimensions on length scales
smaller than 0.1mm \cite{Hoyle}. The minimum bound is given by
\cite{Arkani-Hamed2,Chung}
\begin{equation}\label{brane tension 1}
\lambda \geq (100 \;{\rm GeV})^4.
\end{equation}

With the use of the second transformation in Eqs. (\ref{trans1})
we can express the dimensionless matter density $r _m$ for an
ultra-relativistic gas ($\gamma =4/3$) and with energy-density
$\rho _m=\pi ^2 T^4/15$ as a function of the temperature $T$ as
$r_m=4\pi ^3GT^4/90\alpha _{\phi }^2M_{\phi }^2$. From the
definition of the decay width we obtain the constraint $T<M_{\phi
}$ on the temperature. Therefore, for a universe in which the
normal matter is in the form of radiation only, this condition can
be reformulated in a form of a constraint on the dimensionless
matter density $r_m$, given by
\begin{equation}\label{condn}
r_{m}<\frac{4\pi ^{3}GM_{\phi }^{2}}{90\alpha _{\phi }^{2}}.
\end{equation}
On the other hand from Eqs. (\ref{f3}) and (\ref{f4}) we obtain
the condition $\omega r_{m}^{1/2}<1$, which is consistent with Eq.
(\ref{condn}).

\section{Extra-dimensional effects dominated brane reheating}\label{max bound of C_1}

At the end of the inflation, there is no ordinary matter present
in the universe. Even in the first early stages of the reheating,
the scalar field dominates the matter energy densities, $\rho_\phi
\gg \rho_m,\rho _{DM}$. Due to the high energy of the scalar
field, the extra-dimensional effects due to the presence of the
fifth dimension are important at this stage,and the quadratic
terms in the evolution equations dominate the cosmological
dynamics. Therefore one can neglect the linear term in Eq.
(\ref{f1}) with respect to the quadratic term. Hence,  the
extra-dimensional effects are dominant during the reheating era.
Moreover, we assume that the condition $T<<M_{\phi }$ is satisfied
during the entire reheating process. This condition implies
$\omega r_m^{2\left(\gamma -1\right)/\gamma }<<1$. Then the
equations describing the reheating process with bulk-brane energy
transfer are given by
\begin{equation}
\frac{1}{a}\frac{da}{d\tau }=\eta \left( r_{m}+r_{DM}+r_{\phi
}\right) , \label{g1}
\end{equation}
\begin{equation}
\frac{dr_{m}}{d\tau }+3\gamma \frac{1}{a}\frac{da}{d\tau
}r_{m}=r_{\phi }, \label{g3}
\end{equation}
\begin{equation}
\frac{dr_{\phi }}{d\tau }+6\frac{1}{a}\frac{da}{d\tau }r_{\phi
}=-r_{\phi }. \label{g4}
\end{equation}
\begin{equation}
\frac{dr_{DM}}{d\tau }+3w\frac{1}{a}\frac{da}{d\tau
}r_{DM}=\varepsilon \left( \frac{a_{0}}{a}\right)
^{3w}\frac{1}{a}\frac{da}{d\tau },  \label{g2}
\end{equation}

Eq. (\ref{g4}) can be immediately integrated, and with the use of
the initial conditions we obtain the evolution of the energy
density of the scalar field during reheating as
\begin{equation}\label{rf}
r_{\phi }\left( \tau \right) =r_{\phi 0}\left(
\frac{a_{0}}{a}\right) ^{6}e^{-\tau }.
\end{equation}

Eq. (\ref{g2}) gives the density of the dark matter on the brane
in the form
\begin{equation}\label{rd}
r_{DM}\left( \tau \right) =\varepsilon \left(
\frac{a_{0}}{a}\right) ^{3w}\ln \left( \frac{a}{a_{0}}\right) .
\end{equation}

The evolution of the matter energy density may be represented as
\begin{equation}\label{rm}
r_{m}(\tau )=r_{\phi 0}a_{0}^{6-3\gamma }\left(
\frac{a_{0}}{a}\right) ^{3\gamma }\int _0^{\tau }a^{3\gamma
-6}e^{-\tau }d\tau .
\end{equation}

\subsection{The stiff matter case}

The value $\gamma =2$ corresponds to the case of the so-called
stiff matter fluid, with equation of state given by $p_{m}=\rho
_{m}$.  In this case the energy density of the matter is given by
\begin{equation}
r_{m}(\tau )=r_{\phi 0}\left( 1-e^{-\tau }\right) \left( \frac{a_{0}}{a}%
\right) ^{6},\gamma =2.
\end{equation}

The sum of the energy densities of the matter and scalar field
satisfies the equation
\begin{equation}
\frac{d}{d\tau }\left( r_{m}+r_{\phi }\right) =-\left( 3\gamma
r_{m}+6r_{\phi }\right) \frac{1}{a}\frac{da}{d\tau }.
\end{equation}

For $\gamma =2$ we obtain
\begin{equation}
r_{m}+r_{\phi }=r_{\phi 0}\left( \frac{a_{0}}{a}\right) ^{6}.
\end{equation}

The scale factor of the universe can be obtained from the general
equation
\begin{eqnarray}
\frac{1}{a}\frac{da}{d\tau }&=&\eta \left\{ r_{\phi
0}a_{0}^{6-3\gamma }\left( \frac{a_{0}}{a}\right) ^{3\gamma
}\right.\left[ \int a^{3\gamma -6}e^{-\tau }d\tau -\beta \left(
a_{0}\right) \right] + \nonumber\\
&&\varepsilon \left( \frac{a_{0}}{a}\right) ^{3w}\ln \left( \frac{a}{a_{0}}%
\right) +\left.r_{\phi 0}\left( \frac{a_{0}}{a}\right)
^{6}e^{-\tau }\right\} .
\end{eqnarray}

In the case in which both the dark matter and the usual matter are
in the ultra-relativistic state, $\gamma =w=2$, and the equation
describing the scale factor dynamics is given by
\begin{equation}
a^{5}\frac{da}{d\tau }=\eta a_{0}^{6}\left[ r_{\phi 0}+\varepsilon
\ln \left( \frac{a}{a_{0}}\right) \right] ,\gamma =w=2.
\end{equation}

In this case the time evolution of the universe during the
extra-dimensions dominated phase of the reheating era can be
expressed as
\begin{equation}
\tau =\frac{1}{\varepsilon \eta }\exp \left( -\frac{6r_{\phi
0}}{\varepsilon
}\right) {\rm Ei}\left[ \frac{6r_{\phi 0}}{\varepsilon }+6\ln \left( \frac{a%
}{a_{0}}\right) \right] ,
\end{equation}
where ${\rm Ei}\left( z\right) =-\int_{-z}^{\infty }e^{-t}dt/t$ is
the exponential integral function, and where the principal value
of the integral must be taken. 
In a power series representation
the dimensionless time parameter $\tau $ can be obtained in terms
of the scale factor $a$ as
\begin{equation}
\tau \approx \frac{1}{\varepsilon \eta }\left(
\frac{a}{a_{0}}\right)
^{6}\left\{ \frac{1}{6r_{\phi 0}/\varepsilon +6\ln \left( a/a_{0}\right) }+%
\left[ \frac{1}{6r_{\phi 0}/\varepsilon +6\ln \left(
a/a_{0}\right) }\right] ^{2}+...\right\} .
\end{equation}

\subsection{The radiation dominated phase}

In order to obtain a clear physical picture of the reheating
process during the extra-dimensional effects dominated era one
must study the evolution equations Eqs. (\ref{g1}) - (\ref{g2})
numerically. In order to do this we assume that the normal matter
is always generated from the inflaton field in the form of
ultra-relativistic particles, leading to a very hot and high
density initial state of the universe, an assumption which is
consistent with the standard hot Big-Bang model. Therefore we take
$\gamma =4/3$. As for the dark matter, it can be injected from the
bulk both as a relativistic fluid, with $w=4/3$, or as a
non-relativistic, pressureless fluid, with $w=1$. The variation of
the radiation energy density is represented, for both equations of
state of the dark matter, for fixed values of the parameters
$\varepsilon $ and $\omega $ and for different values of the
parameter $\eta $ in Fig. 1.

\begin{figure}[!ht]
\centering
\includegraphics[width=.48\textwidth]{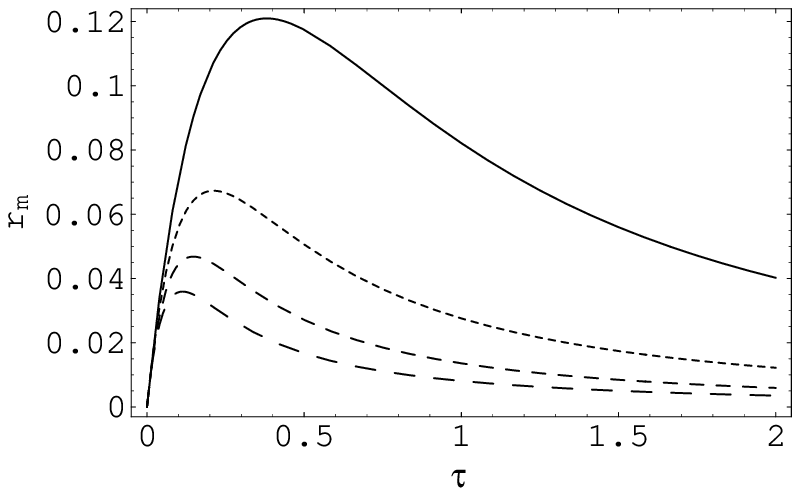}
\includegraphics[width=.48\textwidth]{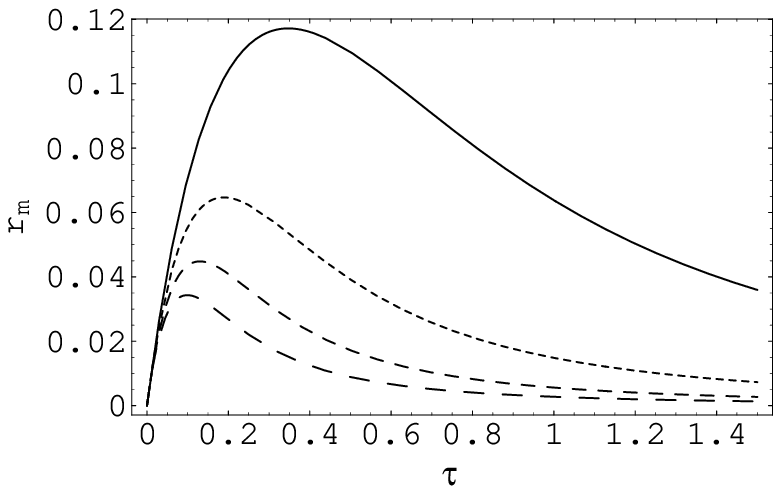}
\caption{Variation of the energy density of the radiation ($\gamma
=4/3$) during the extra-dimensional effects dominated reheating
period, with ultra-relativistic dark matter injection from the
bulk ($w=4/3$) (figure on the left) and non-relativistic
pressureless dark matter injection ($w=1$) (figure on the right),
respectively, for $\omega =0.5$, $\varepsilon =6$ and different
values of $\eta $: $\eta =0.5$ (solid curve), $\eta =1$ (dotted
curve), $\eta =1.5$ (short dashed curve) and $\eta =2$ (long
dashed curve). The initial values of the scalar field density and
of the scale factor are $r_{\phi 0}=1$ and $a_0=10^{-3}$,
respectively.} \label{fig1}
\end{figure}

The variation of the dark matter energy density is represented in
Fig. 2.

\begin{figure}[!ht]
\centering
\includegraphics[width=.48\textwidth]{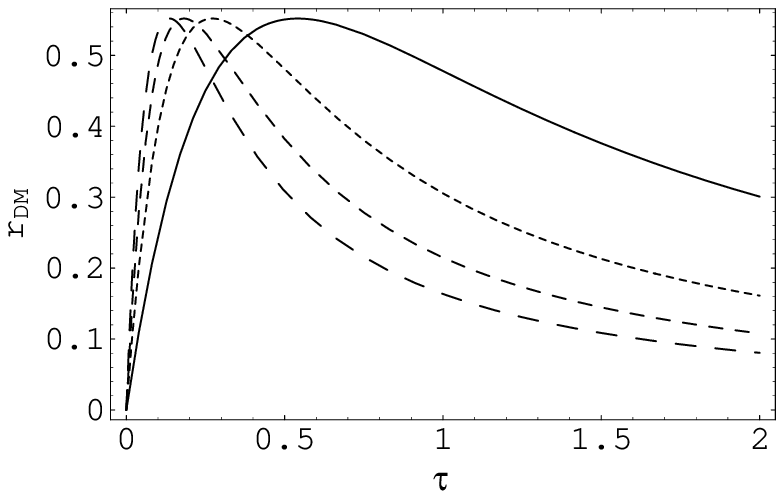}
\includegraphics[width=.48\textwidth]{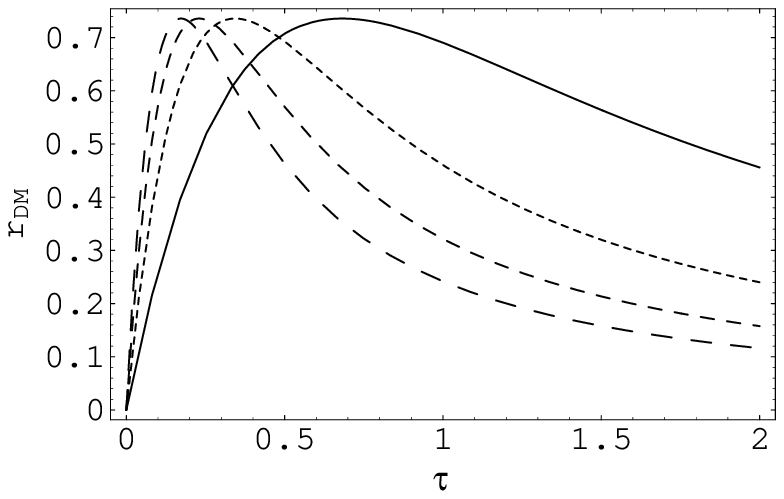}
\caption{Variation of the energy density of the ultra-relativistic
dark matter ($w=4/3$) (figure on the left) and of the pressureless
non-relativistic dark matter ($w=1$) (figure on the right),
respectively, during the extra-dimensional effects dominated
reheating period, for $\omega =0.5$, $\varepsilon =6$ and
different values of $\eta $: $\eta =0.5$ (solid curve), $\eta =1$
(dotted curve), $\eta =1.5$ (short dashed curve) and $\eta =2$
(long dashed curve). The initial values of the scalar field
density and of the scale factor are $r_{\phi 0}=1$ and
$a_0=10^{-3}$, respectively. The normal matter is in the form of
radiation with $\gamma =4/3$.} \label{fig2}
\end{figure}

The variations of the scalar field energy densities and of the
scale factors in these two situations are represented in Figs. 3
and 4, respectively.

\begin{figure}[!ht]
\centering
\includegraphics[width=.48\textwidth]{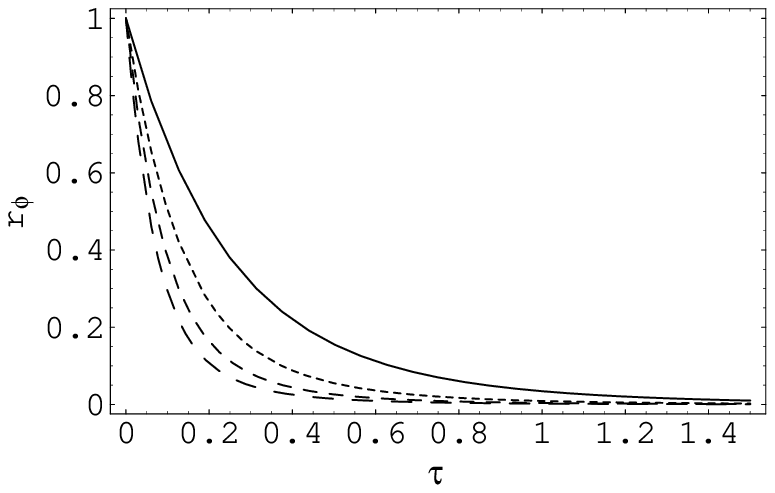}
\includegraphics[width=.48\textwidth]{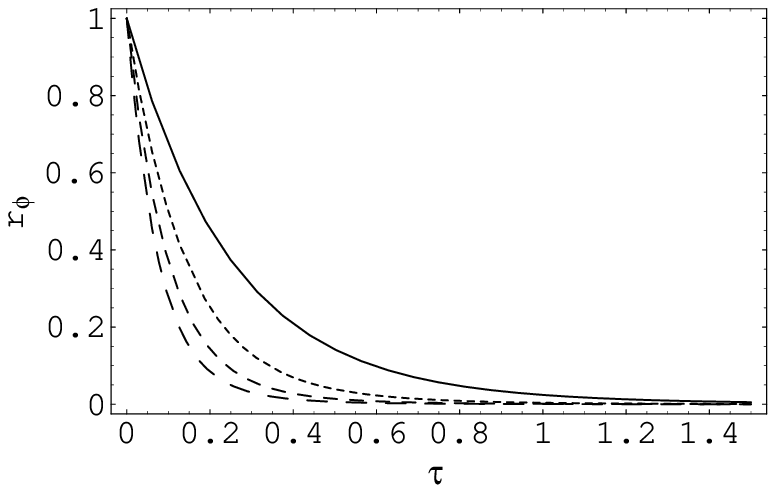}
\caption{Time variation of the energy density of the inflaton
field during the extra-dimensional effects dominated reheating
period, with ultra-relativistic dark matter injection from the
bulk ($w=4/3$) (figure on the left) and non-relativistic
pressureless dark matter injection ($w=1$) (figure on the right),
respectively, for $\omega =0.5$, $\varepsilon =6$ and different
values of $\eta $: $\eta =0.5$ (solid curve), $\eta =1$ (dotted
curve), $\eta =1.5$ (short dashed curve) and $\eta =2$ (long
dashed curve). The initial values of the scalar field density and
of the scale factor are $r_{\phi 0}=1$ and $a_0=10^{-3}$,
respectively. The normal matter is in the form of radiation with
$\gamma =4/3$.} \label{fig3}
\end{figure}

\begin{figure}[!ht]
\centering
\includegraphics[width=.48\textwidth]{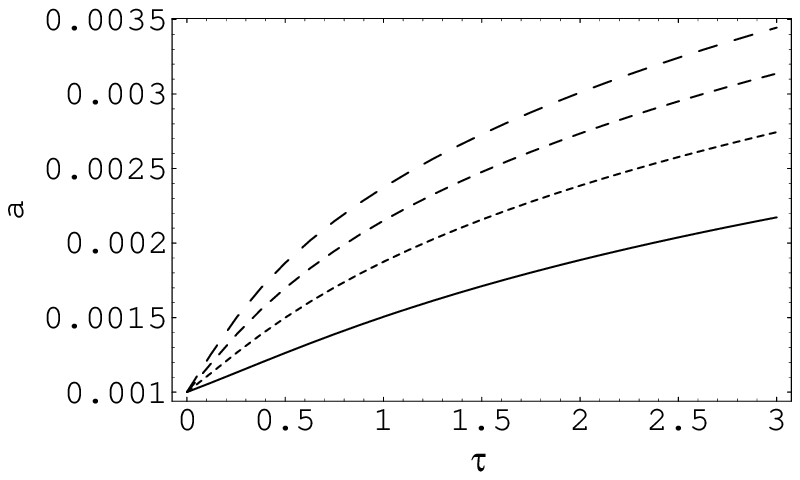}
\includegraphics[width=.48\textwidth]{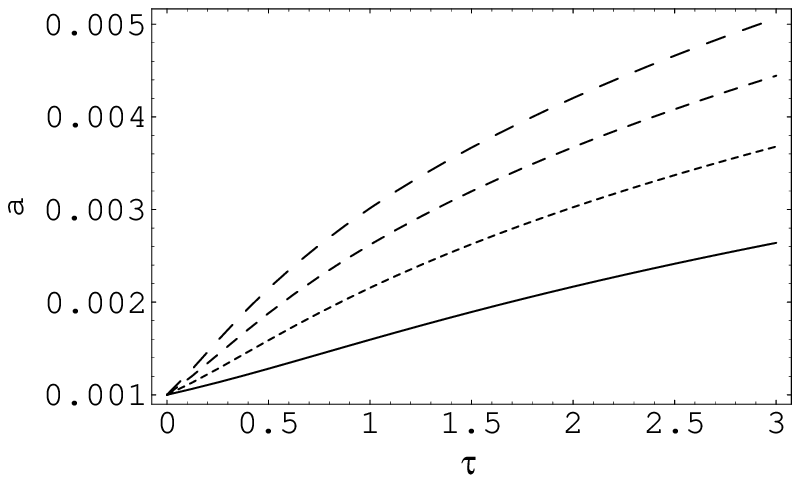}
\caption{Time variation of the scale factor of the universe during
the extra-dimensional effects dominated reheating period, with
ultra-relativistic dark matter injection from the bulk ($w=4/3$)
(figure on the left) and non-relativistic pressureless dark matter
injection ($w=1$) (figure on the right), respectively, for $\omega
=0.5$, $\varepsilon =6$ and different values of $\eta $: $\eta
=0.5$ (solid curve), $\eta =1$ (dotted curve), $\eta =1.5$ (short
dashed curve) and $\eta =2$ (long dashed curve). The initial
values of the scalar field density and of the scale factor are
$r_{\phi 0}=1$ and $a_0=10^{-3}$, respectively. The normal matter
is in the form of radiation with $\gamma =4/3$.} \label{fig4}
\end{figure}

As one can see from the Figs. 1 and 2, during the reheating period
the energy of the scalar field is transferred to the normal matter
(radiation). The energy density of the matter increases from zero
to a maximum value, and, after reaching the maximum, it starts to
decrease, due to the cosmological expansion. During the same
period there is an injection of dark matter from the bulk, the
energy density of the dark matter increasing from an initial zero
value to a maximum value, which depends on the equation of state
of the dark matter and on the parameter $\eta $, describing the
magnitude of the extra-dimensional effects. During the reheating
process the energy density of the scalar field decreases to zero.

\section{General evolution of the brane universe during the reheating
period}

After reaching its maximum value during the reheating period, due
to the cosmological expansion, the energy density of the matter
(in all its forms) decreases in time and for time intervals
$t>t_{cr}$ the condition $\rho _{total}/\lambda <<1$ is satisfied.
In this case the quadratic contributions to the field equations
can be neglected, and the universe is described by the standard
general relativistic cosmological models. In the dimensionless
formulation of the field equations the general relativistic limit
corresponds to $\eta \rightarrow \infty $. In the general case,
when all terms are considered in the field equations, the
evolutions of the matter and of the scalar field is given
generally by Eqs. (\ref{rf}) - (\ref{rm}), respectively. However,
the linear term in energy density modifies the
expansion rate of the universe, which is described by Eq. (\ref{f1}%
).

From the mathematical structure of the field equations one can
derive some general properties of the model, and of the
cosmological evolution during the reheating. The normal matter and
the dark matter  reach a maximum value at a moment $\tau =\tau
_{\max }$, the maximum values of the densities being given by the
solutions of the equations $dr_{m}/d\tau =0$ and $r_{DM}/d\tau
=0$, respectively. In the case of the dark matter the maximum
value is given by
\begin{equation}
r_{DM}^{(\max )}=\left.\frac{\varepsilon }{3w}\left(
\frac{a_{0}}{a}\right) ^{3w}\right| _{\tau =\tau _{\max }}.
\end{equation}

The maximum value of the dark matter density is determined by the
injection rate $\varepsilon $, the cosmological dynamics,
determining value of the scale factor at $\tau _{\max }$, and by
the equation of state of the dark matter. By assuming that $\omega
r_{m}^{2\left( \gamma -1\right) /\gamma }<<1 $, the maximum value
of the normal matter density can be obtained from the algebraic
equation
\begin{equation}\label{rmm}
r_{m}^{(\max )}=\left.\frac{r_{\phi 0}}{3\gamma h\left( \tau \right) }\left( \frac{%
a_{0}}{a}\right) ^{6}e^{-\tau }\right| _{\tau =\tau _{\max }}.
\end{equation}

In the case of the extra-dimensional effects dominated evolution,
$h\left(
\tau \right) =\eta \left( r_{m}+r_{DM}+r_{\phi }\right) $, and Eq. (\ref{rmm}%
) gives
\begin{equation}
r_{m}^{(\max )}=\frac{r_{\phi }\left( \tau _{\max }\right) +r_{DM}^{(\max )}%
}{2}\left\{ \sqrt{1+\frac{4r_{\phi }\left( \tau _{\max }\right)
}{3\gamma \eta \left[ r_{\phi }\left( \tau _{\max }\right)
+r_{DM}^{(\max )}\right] ^{2}}}-1\right\} ,
\end{equation}
which can be approximated as
\begin{equation}
r_{m}^{(\max )}\approx \frac{r_{\phi }\left( \tau _{\max }\right)
}{3\gamma
\eta \left[ r_{\phi }\left( \tau _{\max }\right) +r_{DM}^{(\max )}\right] }=%
\frac{1}{3\gamma \eta }\left[ 1+\frac{r_{DM}^{(\max )}}{r_{\phi
}\left( \tau _{\max }\right) }\right] .
\end{equation}

The maximum value of the matter energy density is inverse
proportional to the parameter $\eta =\sqrt{3/4\pi G\lambda }\alpha
_{\phi }M_{\phi }$. In the case of a fixed value of $\alpha _{\phi
}M_{\phi }$, a large value of $\eta $, corresponding, for example,
 to a small brane tension $\lambda $, would determine a decrease of the
maximum matter density, while small values of $\eta $ (large brane
tensions) would increase the amount of matter produced during the
reheating phase. In terms of the scale factor the maximum matter
density can be expressed as
\begin{equation}
r_{m}^{(\max )}\approx \frac{1}{3\gamma \eta }\left\{ 1+\frac{\varepsilon }{%
3wr_{\phi 0}}\left[ \frac{a_{0}}{a\left( \tau _{\max }\right)
}\right] ^{3w-6}e^{\tau _{\max }}\right\} .
\end{equation}

Generally, the ratio $u$ of the dark matter and normal matter
energy density can be represented, as a function of the scale
factor, as
\begin{equation}
u=\frac{r_{DM}\left( \tau \right) }{r_{m}\left( \tau \right) }=\frac{%
\varepsilon }{r_{\phi 0}}a_{0}^{3\gamma -6}\left(
\frac{a_{0}}{a}\right) ^{3w-3\gamma }\frac{\ln \left(
a/a_{0}\right) }{\int _0^\tau a^{3\gamma -6}e^{-\tau }d\tau }.
\end{equation}



In the small time limit the energy density of the baryonic and of
the dark matter is negligibly small with respect to the energy
density of the scalar field, which in the very early stages of the
reheating is given by Eq. (\ref{rf}). Therefore in the limit of
$\tau <<1$, the evolution of the scale factor is obtained as
$a\approx a_0\left(6\eta r_{\phi 0}\right)^{1/6}\tau ^{1/6}$.
Hence the small time behavior of $u$ is described by the equation
\begin{equation}
u\approx \frac{\epsilon \left(6\eta \right)^{1/6}}{4r_{\phi
0}^{2/3}}\frac{\ln\left(6\eta r_{\phi 0}\tau \right)}{\tau
^{2/3}}, \tau <<1,
\end{equation}

Since in the limit of large time the function $a^{3\gamma
-6}e^{-\tau }$ tends to zero, it follows that the value of the
integral $\int _0^\tau a^{3\gamma -6}e^{-\tau }d\tau $ can be
written as a function of the initial value of the scale factor
$a_0$, so that $\int _0^\tau a^{3\gamma -6}e^{-\tau }d\tau \approx
\beta \left(a_0\right)$.

Therefore in the large time limit the ratio of the densities of
the dark matter and of the baryonic matter can be written as
\begin{equation}
u=\frac{%
\varepsilon a_{0}^{3\gamma -6}}{r_{\phi 0} \beta \left(
a_{0}\right)  }\left( \frac{a_{0}}{a}\right) ^{3w-3\gamma }\ln
\left( \frac{a}{a_{0}}\right),\tau >>1.
\end{equation}

Precision cosmological observations have shown that $u$ has a
constant numerical value, at a high degree of accuracy. Since in
the present day universe the baryonic matter is non-relativistic,
with $\gamma =1$, the constancy of $u$ requires for the dark
matter to also be in a non-relativistic (cold) phase, with $w=1$.
The magnitude of the dark matter/baryonic matter ratio depends on
the initial conditions of the universe during the reheating
period, namely, the dark matter injection rate, the energy density
of the scalar field and the value of the scale factor. On the
other hand, the ratio of the densities of the dark matter and of
the normal matter is not an {\it absolute constant} during the
cosmological evolution, but it is a very slowly, logarithmically
increasing function of time.

The results of the numerical integration of the Eqs. (\ref{f1}) -
(\ref{f2}) are represented in Figs. 5-8, respectively. Fig. 5
presents the variation of the normal matter energy density during
reheating, with the linear terms in the density also included, for
the two equations of state of the dark matter. The variation of
the dark matter energy density is represented in Fig. 6.

\begin{figure}[!ht]
\centering
\includegraphics[width=.48\textwidth]{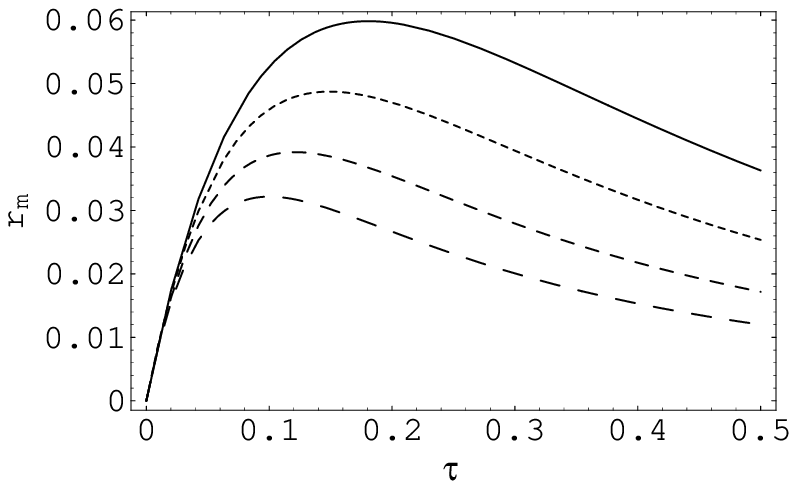}
\includegraphics[width=.48\textwidth]{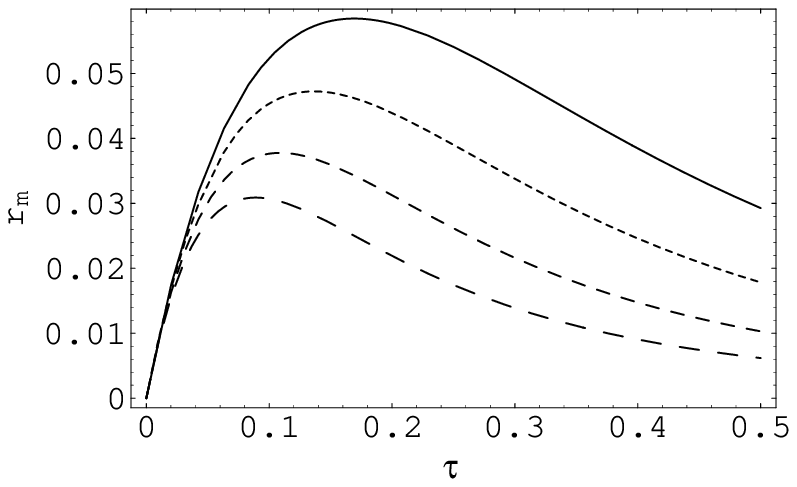}
\caption{Variation of the energy density of the photon gas
($\gamma =4/3$) during the reheating period, with
ultra-relativistic dark matter injection from the bulk ($w=4/3$)
(figure on the left) and non-relativistic pressureless dark matter
injection ($w=1$) (figure on the right), respectively, for $\omega
=0.5$, $\varepsilon =6$ and different values of $\eta $: $\eta
=0.5$ (solid curve), $\eta =1$ (dotted curve), $\eta =1.5$ (short
dashed curve) and $\eta =2$ (long dashed curve). The initial
values of the scalar field density and of the scale factor are
$r_{\phi 0}=1$ and $a_0=10^{-3}$, respectively.} \label{fig5}
\end{figure}

\begin{figure}[!ht]
\centering
\includegraphics[width=.48\textwidth]{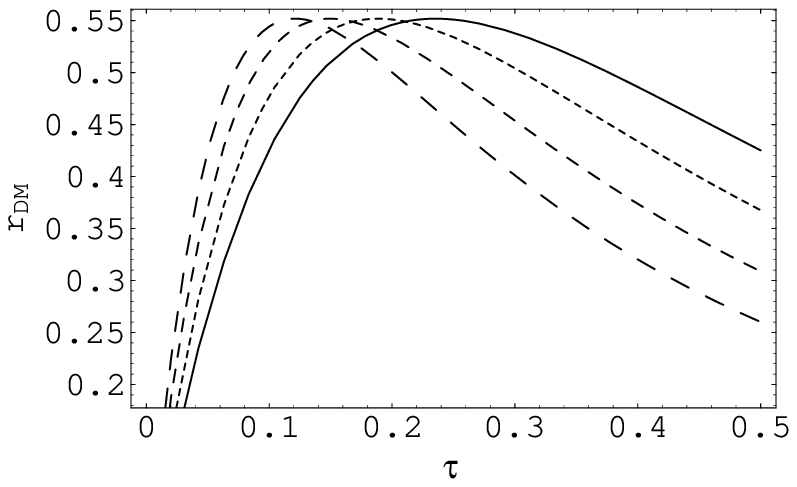}
\includegraphics[width=.48\textwidth]{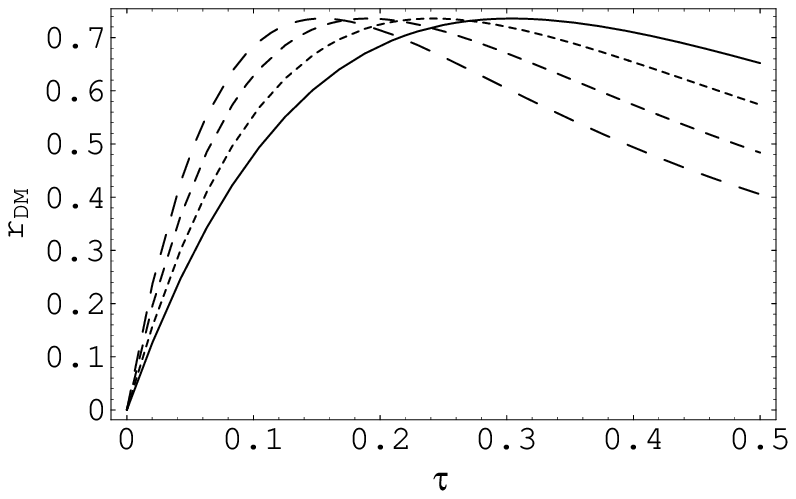}
\caption{Variation of the energy density of the ultra-relativistic
dark matter ($w=4/3$) (figure on the left) and of the pressureless
non-relativistic dark matter ($w=1$) (figure on the right) during
the reheating period, for $\omega =0.5$, $\varepsilon =6$ and
different values of $\eta $: $\eta =0.5$ (solid curve), $\eta =1$
(dotted curve), $\eta =1.5$ (short dashed curve) and $\eta =2$
(long dashed curve). The initial values of the scalar field
density and of the scale factor are $r_{\phi 0}=1$ and
$a_0=10^{-3}$, respectively. The normal matter is in the form of
radiation with $\gamma =4/3$.} \label{fig6}
\end{figure}

The variations of the scalar field energy densities and of the
scale factors for the two equations of state are represented in
Figs. 7 and 8, respectively.

\begin{figure}[!ht]
\centering
\includegraphics[width=.48\textwidth]{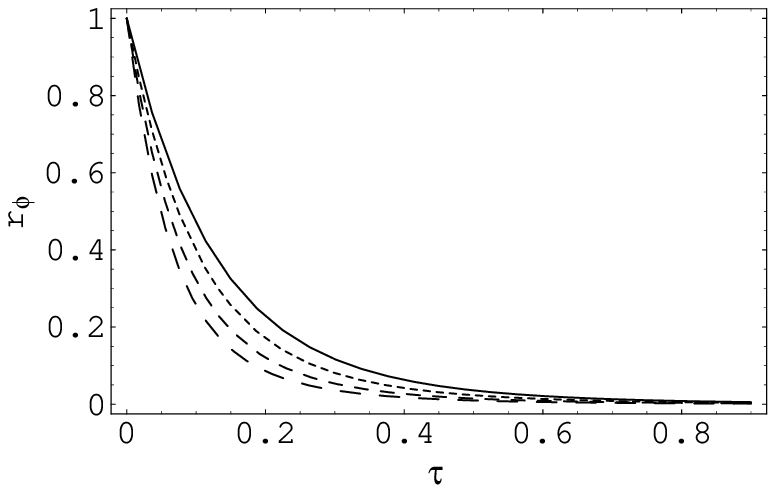}
\includegraphics[width=.48\textwidth]{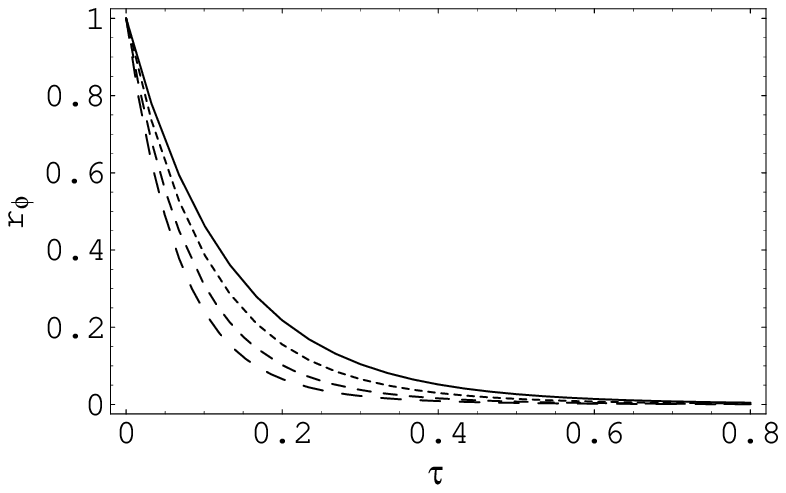}
\caption{Time variation of the energy density of the inflaton
field during the reheating period, with ultra-relativistic dark
matter injection from the bulk ($w=4/3$) (figure on the left) and
non-relativistic pressureless dark matter injection ($w=1$)
(figure on the right), respectively, for $\omega =0.5$,
$\varepsilon =6$ and different values of $\eta $: $\eta =0.5$
(solid curve), $\eta =1$ (dotted curve), $\eta =1.5$ (short dashed
curve) and $\eta =2$ (long dashed curve). The initial values of
the scalar field density and of the scale factor are $r_{\phi
0}=1$ and $a_0=10^{-3}$, respectively. The normal matter is in the
form of radiation with $\gamma =4/3$.} \label{fig7}
\end{figure}

\begin{figure}[!ht]
\centering
\includegraphics[width=.48\textwidth]{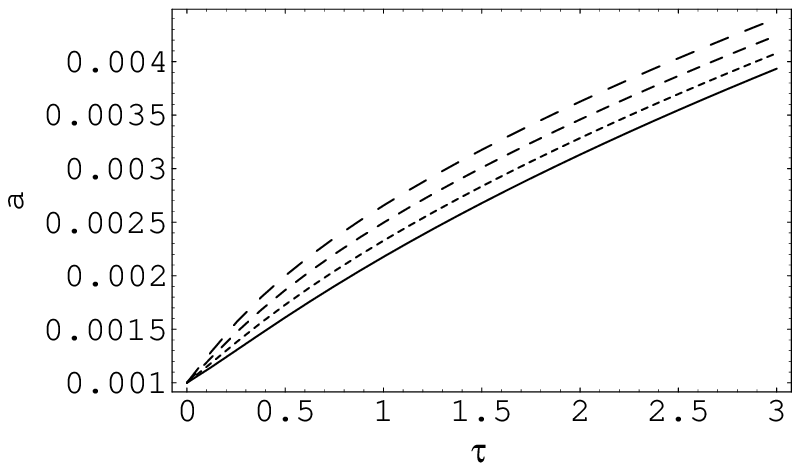}
\includegraphics[width=.48\textwidth]{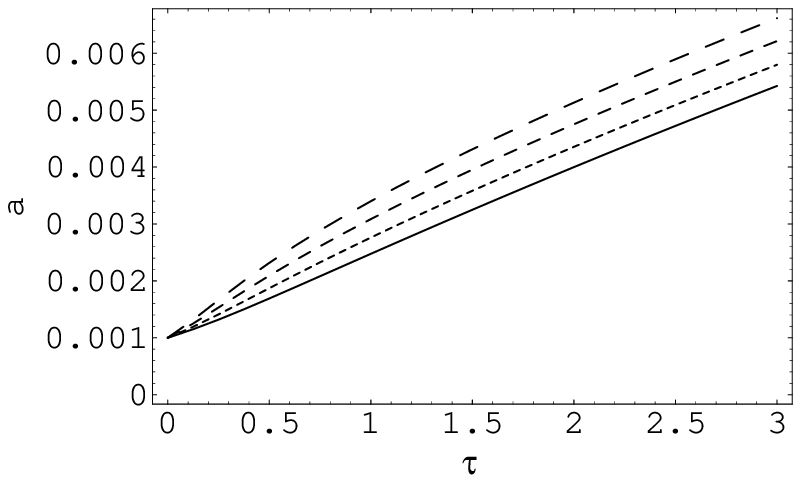}
\caption{Time variation of the scale factor of the universe during
the reheating period, with ultra-relativistic dark matter
injection from the bulk ($w=4/3$) (figure on the left) and
non-relativistic pressureless dark matter injection ($w=1$)
(figure on the right), respectively, for $\omega =0.5$,
$\varepsilon =6$ and different values of $\eta $: $\eta =0.5$
(solid curve), $\eta =1$ (dotted curve), $\eta =1.5$ (short dashed
curve) and $\eta =2$ (long dashed curve). The initial values of
the scalar field density and of the scale factor are $r_{\phi
0}=1$ and $a_0=10^{-3}$, respectively. The normal matter is in the
form of radiation with $\gamma =4/3$.} \label{fig8}
\end{figure}

As one can see from the figures, the inclusion of the linear
matter terms into the evolution equations leads to an overall
decrease of the numerical values of the cosmological parameters
during the reheating period. If in the case of the
extra-dimensional effects dominated dynamics the maximum values of
the matter energy density are of the orders of $r_m\approx 0.12$,
in the general case and for the same value of the physical
parameters this value is $r_m\approx 0.06$. However, the maximum
value of the energy density of the dark matter is not
significantly influenced by the change in the cosmological
dynamics, since it is essentially determined by the injection
process from the bulk. On the other hand the decay of the inflaton
field is more rapid when the full dynamics on the brane is taken
into account.

It is easy to show that the ratio of the dark matter and normal
matter energy density $u$ satisfies the following evolution
equation
\begin{equation}
\frac{du}{d\tau }=\left[ \left( 3\gamma -3w\right) h(\tau )-\frac{r_{\phi }}{%
r_{m}}\sqrt{1-\omega r_{m}^{2\left( \gamma -1\right) /\gamma
}}\right] u+\varepsilon \left( \frac{a_{0}}{a}\right)
^{3w}\frac{h(\tau )}{r_{m}}.
\end{equation}

The maximum value of $u$, $u_{\max }=u\left( \tau _{\max }\right)
$ is given by the solution of the equation $\left.du/d\tau \right|
_{\tau =\tau _{\max }}=0$, giving for the maximum value of $u$ the
expression
\begin{equation}
u_{\max }=\left.\frac{\varepsilon \left( a_{0}/a\right) ^{3w}h(\tau )}{r_{\phi }%
\sqrt{1-\omega r_{m}^{2\left( \gamma -1\right) /\gamma }}-\left(
3\gamma -3w\right) h(\tau )r_{m}}\right| _{\tau =\tau _{\max }}.
\end{equation}

In the limit of large times, when $r_{\phi }\approx 0$, and
assuming that
the dark matter obeys the same equation of state as the normal matter, $%
w=\gamma $, $u$ satisfies the equation
\begin{equation}
\frac{du}{d\tau }=\varepsilon \left( \frac{a_{0}}{a}\right) ^{3w}\frac{%
h(\tau )}{r_{m}}.
\end{equation}

For non-relativistic dark matter and normal matter we obtain the
simple relation
\begin{equation}\label{ratio}
\frac{du}{d\tau }=\varepsilon h(\tau ).
\end{equation}

Thus the variation of the ratio of the dark matter and of the
normal matter is proportional to the Hubble parameter.

To study the time variation of the ratio of the dark and baryonic
matter $u$ we consider that the transition between the
relativistic and non-relativistic state can be modelled as a
smooth transition, with the equations of state of the form
$p_{DM}= w(t)\rho _{DM}$ and $p_m=\gamma (t)\rho _m$,
respectively. In the dimensionless time variable $\tau $ we assume
that the functions $\gamma (\tau )$ and $w(\tau )$ can be given as
\begin{equation}
\gamma (\tau )=\frac{4}{3}\frac{\tau _{o\gamma }}{\tau +\tau
_{0\gamma }}+\frac{\tau }{\tau +\tau _{0\gamma }},
\end{equation}
 and
\begin{equation}
 w(\tau )=\frac{4}{3}\frac{\tau _{ow}}{\tau +\tau
_{0w}}+\frac{\tau }{\tau +\tau _{0w}},
\end{equation}
respectively, where $\tau _{o\gamma }$ and $\tau _{ow}$ are the
critical transitions times for the dark matter and the baryonic
matter. When $\tau <<\tau _{0\gamma },\tau _{0w}$, $\gamma \approx
4/3$, while for $\tau >>\tau _{0\gamma },\tau _{0w}$, $\gamma
\approx 1$. The time variation of $u$ is represented, for two
different numerical values of the critical transitions times of
the dark matter and baryonic matter, respectively, and for
different values of the physical parameters, in Fig. 9.

\begin{figure}[!ht]
\centering
\includegraphics[width=.49\textwidth]{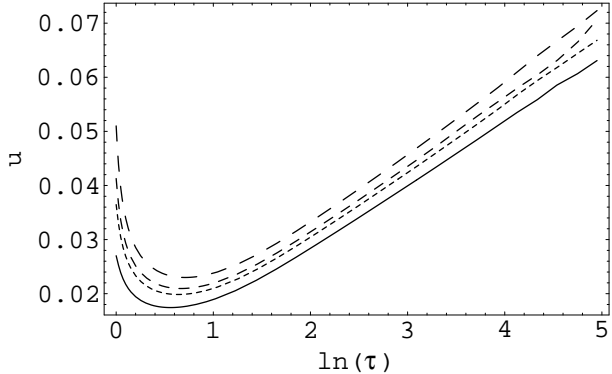}
\caption{Time variation of the ratio $u$ of the dark matter and
normal matter energy densities during the reheating period, by
assuming a smooth transition from the relativistic to the
non-relativistic state, so that $\gamma (\tau )=(4/3)\tau
_{o\gamma }/(\tau +\tau _{0\gamma })+\tau /(\tau +\tau _{0\gamma
})$ and $w (\tau )=(4/3)\tau _{ow }/(\tau +\tau _{0w })\tau /(\tau
+\tau _{0w})$, respectively.  The values of the parameters are
$\omega =0.5$,  $\eta =2$ (solid curve), $\eta =2.5$ (dotted
curve), $\eta =3$ (short dashed curve) and $\eta =3.5$ (long
dashed curve). We have assumed that the transitions from a
non-relativistic to a relativistic phase takes place at the
critical time $\tau _{0\gamma }=\tau _{0w }=10^8$. The initial
values of the scalar field density and of the scale factor are
$r_{\phi 0}=1$ and $a_0=10^{-3}$, respectively, while the dark
matter injection rate is $\varepsilon =0.01$. } \label{fig9}
\end{figure}

\section{Discussions and final remarks}

In the present paper we have considered the reheating of the
universe in the brane world models with bulk-brane energy
exchange. Our model is essentially based on the hypothesis that
the present day dark matter, representing around $25\%$ of the
total matter content of the universe, was transferred onto the
brane during the post-inflationary reheating period, while the
normal matter is the result of the decay of the inflaton field
into a ultra-relativistic photon gas. Hence, after the
inflationary epoch nor dark matter neither normal matter were
present on the brane, and they were created during the reheating.

To study the reheating process we have introduced a set of
dimensionless quantities, which describe the evolution of the
scaled densities in terms of a dimensionless time parameter $\tau
$. By using the results of the numerical simulations for the
rescaled variables one can obtain some constraints on the physical
parameters of the model.

The reheating time $t_{reh}$, the time after which the universe is
reheated to a temperature allowing the standard hot big bang model
to take place is related to the dimensionless time $\tau _{reh}$ by Eqs. (\ref{trans1}) and (%
\ref{trans2}), respectively. These equations allows us to
constrain the mass and the coupling constant of the inflaton field
as
\begin{equation}
\alpha _{\phi }M_{\phi }=\frac{\tau _{reh}}{2t_{reh}}=0.328\tau
_{reh}t_{reh}({\rm s})\times 10^{-24}\;{\rm GeV}.
\end{equation}

Since according the hot standard big-bang scenario reheating must
end at around $t_{reh}=10^{-18}$ s, it follows that $\alpha _{\phi
}M_{\phi }=0.328\tau _{reh}$ keV. In our model and for the typical
values of the physical parameters we have adopted (bulk-brane
injection rate and brane tension), according to the numerical
analysis reheating may definitely take place for values of $\tau
_{reh}$ in the range $0.1\leq \tau _{reh}\leq 1$ (see Figs. 1-2
and 5-6), thus giving $0.0328\;{\rm keV}\leq \alpha _{\phi
}M_{\phi }\leq 0.328\;{\rm keV}$. Values of $\alpha _{\phi
}M_{\phi }$ of the order of 1 keV or higher would require values
of $\tau _{reh}$ of the order of $\tau _{reh}\geq 3$.

The numerical analysis of the field equations, as well as their
qualitative study, shows the existence of an absolute maximum
value of the matter density $\rho _{matter}^{(\max )}$ during the
cosmological evolution of the universe. The maximum value of the
total matter density is given by
\begin{equation}
\rho _{matter}^{(\max )}=\left[ r_{m}^{(\max )}+r_{DM}^{(\max
)}\right] \frac{12\alpha _{\phi }^{2}M_{\phi }^{2}}{8\pi
G}=7.13\left[ r_{m}^{(\max )}+r_{DM}^{(\max )}\right]\left( \alpha
_{\phi }M_{\phi }/{\rm GeV}\right) ^{2}\times 10^{37}\;{\rm
GeV}^{4}.
\end{equation}

By assuming that $r_{m}^{(\max )}+r_{DM}^{(\max )}\approx 0.5-0.7$ and $%
\alpha _{\phi }M_{\phi }\approx 1$ keV, we obtain
\begin{equation}\label{romax}
3.565\times 10^{25}\;{\rm GeV}^{4}\leq \rho _{matter}^{(\max
)}\leq 4. 991\times 10^{25}\;{\rm GeV}^{4}.
\end{equation}

Eq. (\ref{romax}) allows the determination of the reheating temperature $%
T_{reh}$ of the universe, once the equation of state of the matter
is known. Assuming that both normal matter and dark matter are in
a ultra-relativistic state, we obtain
\begin{equation}
T_{reh}\approx 3\times 10^{6}\;{\rm GeV}.
\end{equation}

Several estimations of the numerical value of the reheating
temperature have been obtained by using some observational
constraints.  A strong constraint on the cosmological models comes
from the study of the CMB anisotropy, and requires that the
results do not exceed the observed level of normalized anisotropy
on large scales \cite{infl,Kolb}. By also requiring that the
radiative corrections mediated by the couplings do not spoil the
flatness of the potential limits, it follows that the reheating
temperature must be below the grand unification theory scale,
$T_{reh}<10^{16}$ GeV, which physically means that the grand
unified theory symmetries are not restored and therefore there is
no second phase of production of monopoles \cite{infl,Kolb}. By
assuming that the energy density of the universe is in the form of
relativistic matter with density $\rho _{total}=g_{\ast }\pi
^{2}T^{4}/30$, where $g_{\ast }$ is the effective number of massless degrees of freedom ($%
g_{\ast }=100-1000$), the reheating temperature can be obtained as $%
T_{reh}\approx 0.2\left( 100/g_{\ast }\right) ^{1/4}\sqrt{\Gamma _{tot}m_{Pl}%
}\approx 0.2\left( 100/g_{\ast }\right) ^{1/4}\alpha _{\phi
}^{1/2}\sqrt{M_{\phi }m_{Pl}}$, where $\Gamma _{tot}$ is the total
decay width and $m_{Pl}$ is the Planck mass. From the study of the
CMB anisotropies it follows that $M_{\phi }\approx 10^{-6}m_{Pl}$
\cite{infl,Kolb}.

In the context of supersymmetry models, a stronger constraint may
be obtained from the production rate of gravitinos, the
supersymmetric partners of the gravitons. The gravitinos, having a
long lifetime, of the order of $10^5$ s, may cause serious
problems in cosmology. For a wide range of the gravitino masses in
order to obtain consistency with the nucleosynthesis constraints,
the reheating temperature must be below $10^{9}$ GeV,
$T_{reh}<10^{9}$ GeV \cite{infl,Moroi}. A stronger constraint on
the reheating temperature for the gravitino decay products not to
destroy the light elements synthesized during the Big-Bang
nucleosynthesis was obtained in \cite{OkSe05}, and requires that
$T_{reh}$ must satisfy the condition $T_{reh}\leq 10^{6}-10^{7}$
GeV. Primordial nucleosynthesis requires that the universe is
close to thermal equilibrium at a temperature of around $1-4$ MeV
\cite{reh}. Hence, the reheating temperature predicted by our
model satisfies all the CMB and nucleosynthesis constraints.

By adopting the CMB constraint implying that $M_{\phi }\approx
10^{-6}m_{Pl}\approx 1.22\times 10^{19}$ GeV, we can estimate the
inflaton field coupling constant as $\alpha _{\phi }M_{\phi
}=0.328\tau _{reh}t_{reh}(s)\times 10^{-24}/M_{\phi }=2.69\times
\tau
_{reh}t_{reh}(s)\times 10^{-44}$. For $t_{reh}\approx 10^{-18}$ s we obtain $%
\alpha _{\phi }\approx 2.69\times \tau _{reh}\times 10^{-26}$, or $%
2.69\times 10^{-27}\leq \alpha _{\phi }\leq 2.69\times 10^{-26}$.

The numerical values of the cosmological parameters sensitively
depend on the parameter $\eta $, a dimensionless combination of
the mass and coupling constant of the inflaton field and of the
brane tension, respectively. From Eq. (\ref{trans3}) we obtain the
brane tension $\lambda $ as
\begin{equation}
\lambda =\frac{6\alpha _{\phi }^{2}M_{\phi }^{2}}{8\pi G\eta ^{2}}%
=\frac{3.565}{\eta ^{2}}\left( \alpha _{\phi }M_{\phi }/{\rm
GeV}\right) ^{2}\times 10^{37}\;{\rm GeV}^{4}.
\end{equation}

Since from the results of the numerical integration it follows
that the values of $\eta $ are of the order of $0.5\leq \eta \leq
2$, and by adopting again for $\alpha _{\phi }M_{\phi }$ a value
of 1 keV, we obtain $0.891\times 10^{25}\;{\rm GeV}^4\leq \lambda
\leq 14.26\times 10^{25}\;{\rm GeV}^4$. This value is much larger
than the minimal value $\lambda \geq 10^8\;{\rm GeV}^4$ previously
obtained in \cite{Arkani-Hamed2,Chung}, but it is consistent with
the bound $\lambda \leq 10^{40}$ GeV$^4$ derived in \cite{Mazum}.

In the present model the reheating temperature satisfies with a
very good approximation the relation
\begin{equation}
T_{reh}=\lambda ^{1/4},
\end{equation}
relating the reheating temperature to the brane tension, and
which was first suggested in \cite{Allah}. The universe exits the
extra-dimensions dominated phase, when $\rho _{total}>>\lambda $,
at a density $\rho _{total}\approx \lambda $, which, assuming that
the matter and field content
is formed from ultra-relativistic particles, gives a transition temperature $%
T_{trans}\sim \lambda ^{1/4}$. According to our results,
$T_{trans}\approx T_{reh}$. Therefore, the reheating phase of the
universe essentially took place during the extra-dimensions
dominated cosmological era, when the bulk effects may have played
a major role in the dynamics and cosmological defrosting of the
universe. On the other hand, since the reheating temperature may
be estimated from observations, the exact relation between
$T_{reh}$ and $\lambda $ gives the possibility of the direct
determination of the brane tension from cosmological observations.

The basic feature of the present reheating model is the continuous
energy-matter transfer between the brane and the bulk. Even that
this transfer was maximum during the reheating period, essentially
contributing in establishing the present day cosmological chemical
composition of the universe, a very small rate of change of the
ratio of the dark matter and of the normal matter still occurred
during the later phases of the cosmological evolution. The present
day rate of change of this ratio can be estimated from Eq.
(\ref{ratio}), and is given by
\begin{equation}
\frac{du}{dt}\approx \varepsilon \times H_{0}\approx
3.24\varepsilon \times 10^{-18}\;{\rm s}^{-1},
\end{equation}
where $H_0$ is the present day value of the Hubble parameter.
Since in order to explain the cosmological ratio of the matter
components $\varepsilon $ should have numerical values of the
order of unity, it follows that the time changes of the dark
matter and normal matter ratio are negligible small in the recent
cosmological times.

\section*{Acknowledgments}
This work is supported by the RGC grant HKU 7027/06P of the
Government of the Hong Kong SAR.

\end{document}